\documentclass[fleqn,10pt]{wlscirep}
\usepackage[utf8]{inputenc}
\usepackage[T1]{fontenc}
\usepackage{braket}
\usepackage{color}
\usepackage{units}

\title{Forming individual magnetic biskyrmions by merging two skyrmions in a centrosymmetric nanodisk}

\author[1,*]{B{\"o}rge G{\"o}bel}
\author[2]{J\"urgen Henk}
\author[1,2]{Ingrid Mertig}
\affil[1]{Max-Planck-Institut f\"ur Mikrostrukturphysik, Halle (Saale), 06120, Germany}
\affil[2]{Martin-Luther-Universit\"at Halle-Wittenberg, Institut f\"ur Physik, Halle (Saale), 06099, Germany}

\affil[*]{bgoebel@mpi-halle.mpg.de}

\usepackage{bm}
\renewcommand{\vec}[1]{\boldsymbol{#1}}
\newcommand{\ed}[1]{{\color{black}{#1}}} 
\newcommand{\edb}[1]{{\color{black}{#1}}} 
\newcommand{\edc}[1]{{\color{black}{#1}}} 

\begin{abstract}
\edb{When two magnetic skyrmions -- whirl-like, topologically protected quasiparticles -- form a bound pair, a biskyrmion state with a topological charge of $N_\mathrm{Sk}=\pm 2$ is constituted. Recently, especially the case of two partially overlapping skyrmions has brought about great research interest. Since for its formation} the individual skyrmions \edb{need to} posses opposite in-plane magnetizations, \edb{such a} biskyrmion cannot be stabilized by the Dzyaloshinskii-Moriya-interaction (DMI), which is the interaction that typically stabilizes skyrmions in non-centrosymmetric materials and at interfaces. Here, we show that \edb{these} biskyrmions can be stabilized by the dipole-dipole interaction in centrosymmetric materials in which the DMI is forbidden. Analytical considerations \edc{indicate that the bound state of a biskyrmion is energetically preferable over two individual skyrmions}. As a result, when starting from two skyrmions \edb{in a micromagnetic simulation}, a biskyrmion is formed upon relaxation. We propose a scheme that allows to control this biskyrmion formation in nanodisks and analyze the individual steps.
\end{abstract}
\begin{document}

\flushbottom
\maketitle
\thispagestyle{empty}

\section*{Introduction}

Magnetic skyrmions~\cite{bogdanov1989thermodynamically,muhlbauer2009skyrmion, nagaosa2013topological} are whirl-like magnetic quasiparticles on the sub-micrometer scale, first observed in MnSi~\cite{muhlbauer2009skyrmion} as periodic arrays (so-called skyrmion crystals or lattices). 
Later, also isolated skyrmions have been found in ferromagnetic Fe$_{0.5}$Co$_{0.5}$Si films~\cite{yu2010real}. Each skyrmion is characterized by a topological charge 
\begin{align}
N_\mathrm{Sk}= \frac{1}{4\pi} \int \vec{m}(\vec{r}) \cdot \left[ \frac{\partial \vec{m}(\vec{r})}{\partial x}  \times  \frac{\partial \vec{m}(\vec{r})}{\partial y}  \right]\,\mathrm{d}r^{2}\label{eq:nsk}
\end{align}
of $\pm 1$ [$\vec{m}(\vec{r})$ is the magnetization], which imposes an energy barrier that protects a skyrmion from annihilating to the ferromagnetic groundstate. Their high stability and small sizes make skyrmions candidates for the carriers of information in future storage devices~\cite{parkin2004shiftable,sampaio2013nucleation}. For example, skyrmions can be written and deleted, driven by electric currents and read in thin films~\cite{sampaio2013nucleation,romming2013writing,jiang2017direct, litzius2017skyrmion,maccariello2018electrical,hamamoto2016purely}.

\begin{figure*}[t!]
	\centering
	\includegraphics[width=\textwidth]{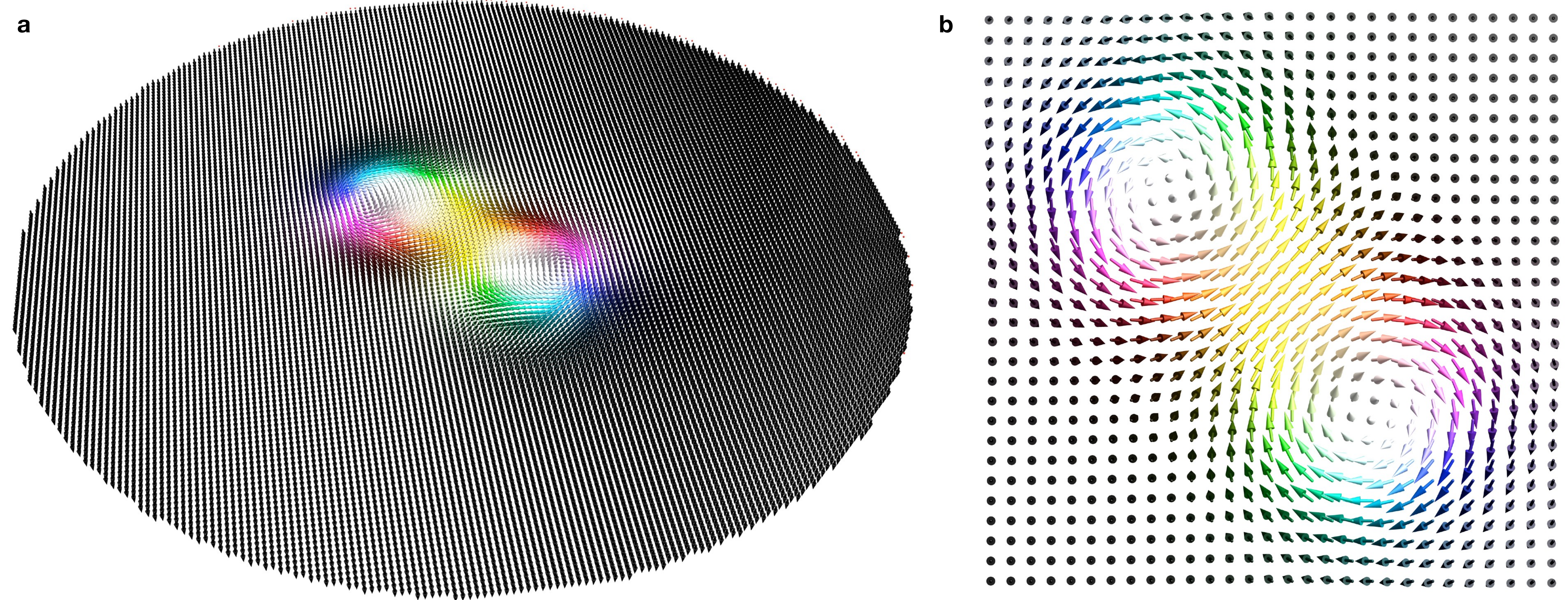}
	\caption{Magnetic biskyrmion. (\textbf{a}) Spin texture of a magnetic biskyrmion in the considered nanodisk resulting from a micromagnetic simulation (also presented in Fig.~\ref{fig:biskyrmion}). The orientation of the individual magnetic moments is represented by a Lorentz transmission electron microscopy (LTEM) color scheme. (\textbf{b}) Close-up of the center region of (a). The biskyrmion consists of two overlapping skyrmions with helicities $\pm \pi/2$, respectively. In the two panels only every fourth or sixteenth magnetic moment from the simulation is displayed, respectively, for better visibility.}
	\label{fig:overview}
\end{figure*}

\edb{An object which is closely related to the skyrmion is the biskyrmion, a term that describes two skyrmions in a bound state. Like the skyrmion~\cite{skyrme1961non} it was initially proposed in nuclear physics~\cite{biedenharn1986topological,schramm1988calculation}. In the context of magnetism the term `biskyrmion' has been used for two skyrmions in a bilayer quantum Hall system~\cite{hasebe2002grassmannian}, for two asymmetric skyrmions in the cone phase of a chiral helimagnet which exhibit an attractive interaction~\cite{leonov2016three,du2018interaction}, and -- as observed for the first time in 2014~\cite{yu2014biskyrmion} -- a composition of two partially overlapping skyrmions with opposite in-plane magnetizations (Fig.~\ref{fig:overview}b)~\cite{yu2014biskyrmion,wang2016centrosymmetric, peng2017real,zuo2018direct,peng2018multiple,zhang2017skyrmion}. Still, all of these objects are geometrically distinct. In the remainder of this paper the term `biskyrmion' refers always to the latter object: a pair of partially overlapping skyrmions with a helicity difference of $\pi$ that has attracted an enormous research interest since its initial discovery.} 

Although merged, each skyrmion forming the biskyrmion can still be identified and their shifted centers indicate a merely partial overlap. For this reason, each skyrmion contributes with its topological charge of $\pm 1$ to the biskyrmion's topological charge of $N_\mathrm{Sk} = \pm 2$. Periodic arrays of \edb{such} biskyrmions (biskyrmion crystals) have first been observed in the centrosymmetric layered manganite La$_{2-2x}$Sr$_{1+2x}$Mn$_2$O$_7$~\cite{yu2014biskyrmion}; other centrosymmetric biskyrmion hosts followed recently~\cite{wang2016centrosymmetric,peng2017real,zuo2018direct, peng2018multiple}. 

If \emph{individual} biskyrmions could be stabilized, they could be used to store information just like conventional skyrmions. However, up to now, individual biskyrmions have only been predicted in frustrated magnets~\cite{zhang2017skyrmion} --- even the existence of conventional skyrmions in these materials~\cite{okubo2012multiple,leonov2015multiply,gobel2017afmskx} remains to be proven experimentally. These considerations suggest to investigate centrosymmetric materials with dominating dipole-dipole interaction, like in the experimental observations of biskyrmion crystals, to find also individual biskyrmions. \ed{In these materials the constituents of biskyrmions --- skyrmionic bubbles --- have been found already decades ago~\cite{malozemoff2016magnetic,eschenfelder2012magnetic}.}

In this Paper, we propose a scheme for creating biskyrmions in a controlled manner. In a first step, we \edc{derive an attractive skyrmion-skyrmion interaction which is necessary for the formation of a biskyrmion}, using an analytical model. Thereafter, we show by micromagnetic simulations for a nanodisk geometry that two Bloch skyrmions with opposite in-plane magnetizations can be written at opposite sides of a nanodisk. We simulate the resultant motion of such skyrmions upon relaxation and derive their equations of motion analytically. After relaxation we find that both skyrmions have merged to a stable biskyrmion state (Fig.~\ref{fig:overview}). The importance of the skyrmions' helicities for this process is discussed. We summarize our findings and give an outlook.

\section*{Results}

\subsection*{\edc{Analytical superposition of two skyrmions}}
\edc{As presented in the introduction, it is an established fact that non-collinear spin textures like skyrmions can be stabilized by the dipole-dipole interaction~\cite{nagaosa2013topological}. Therefore, it is well conceivable that also two skyrmions with reversed in-plane magnetizations can be stabilized in the same sample by this mechanism. The question is now how these skyrmions interact. As we will show, already considering a short-range approximation of the dipole-dipole interaction leads to the conclusion that the two skyrmions attract each other, thereby allowing to form a biskyrmion. This finding is later confirmed by micromagnetic simulations where the full dipole-dipole interaction is considered and a biskyrmion state is metastabilized. 

To establish the attractive interaction between two skyrmions with opposite in-plane magnetizations, we superpose such two (fixed) skyrmions at a distance of $\Delta r$ analytically}. For simplicity we neglect shape deformations of the skyrmions \edc{(such deformations are accounted for by the micromagnetic simulations later in this paper).} 
The magnetic textures  $\vec{m}(\vec{r}) = (\sin\Theta\cos\Phi, \sin\Theta\sin\Phi, \cos\Theta)$ are imposed onto a two-dimensional square lattice with lattice constant $a$. 

The spherical coordinates of the magnetic moments of an individual skyrmion read $\Phi = \arctan(m \, y/x) + \gamma$ and $\Theta = 2 \arctan(\sqrt{x^2 + y^2} / r_0)$; $r_0$ determines the size of the skyrmion. The helicity $\gamma$ of the skyrmion manifests itself in the in-plane components via $\Phi = m \phi + \gamma$. Here, $\phi$ is the polar angle of the position vector with respect to the skyrmion's center and $m = +1$ is the vorticity, which defines the sense of in-plane spin rotation. In case of a Bloch skyrmion we have $m=1$ and $\gamma = \pm \pi / 2$, i.\,e., the polar angle of the spin orientation $\Phi$ has a fixed offset to the polar angle $\phi$ of the position $\vec{r}$ with respect to the skyrmion's center. The topological charge $N_\mathrm{Sk} = p \cdot m = \pm 1$ of a skyrmion depends on the skyrmion's polarity $p=\pm 1$. In this Paper, we consider skyrmions in a ferromagnetic surrounding, magnetized along $-z$, which gives a polarity and topological charge of  $+1$.

One way to superpose two skyrmions is adding their polar angles and multiplying the arguments of their $\arctan$ function for the azimuthal angle, 
\begin{align}
\label{eq:model}
\begin{split}
\Phi & = \arctan\left(\frac{y-\Delta y/2}{x-\Delta x/2}\right)+\arctan\left(\frac{y+\Delta y/2}{x+\Delta x/2}\right)+(\gamma_d-\alpha), \\
\Theta & = 2\arctan\left[\frac{1}{r_0^2}\sqrt{(x-\Delta x/2)^2+(y-\Delta y /2)^2}\cdot\sqrt{(x+\Delta x/2)^2+(y+\Delta y /2)^2}\right].
\end{split}
\end{align}
The condition $|\vec{m}| = 1$ holds for all $\vec{r}$. Here, $\Delta \vec{r} = (\Delta x, \Delta y, 0) = \Delta r (\cos\alpha, \sin\alpha, 0)$ is the displacement vector of the two skyrmions with helicities $\gamma_d$ and $\gamma_d+\pi$. For an inter-skyrmion distance of $\Delta r \gg r_0$ we find two isolated skyrmions, for $\Delta r \approx 2r_0$ the solution is a biskyrmion state with two skyrmions overlapping partially, and for $\Delta r = 0$ a higher-order skyrmion is formed. In any case, the total topological charge is $N_\mathrm{Sk} = +2$.

\subsection*{\edc{Attractive skyrmion-skyrmion interaction}}
\edc{We clarify now which term in the lattice Hamiltonian causes an attractive interaction of the two skyrmions, what can lead to the formation of a biskyrmion.} The exchange interaction $H_\mathrm{ex} = -\frac{1}{2} \sum_{ij} J_{ij}\vec{m}_i \cdot \vec{m}_j$ is independent of $\Delta r$ and independent of the skyrmions' sizes and shapes. Hence, the exchange interaction alone will not stabilize a biskyrmion. This finding is in line with the fact that even conventional skyrmions cannot be stabilized by a bare exchange interaction, unless further (frustrated) exchange interactions are considered; then, the interaction constants $J_{ij}$ have different signs for nearest and second-nearest neighbor spins. The same requirement holds for biskyrmions\cite{zhang2017skyrmion}.

\edc{In most materials,} skyrmions are stabilized by the Dzyaloshinskii-Moriya interaction (DMI)~\cite{dzyaloshinsky1958thermodynamic, moriya1960anisotropic} $H_\mathrm{DMI} = \frac{1}{2}\sum_{ij} \vec{D}_{ij} \cdot (\vec{m}_i \times \vec{m}_j)$ that originates from spin-orbit coupling and a broken inversion symmetry. The DMI vectors $\vec{D}_{ij} = -\vec{D}_{ji}$ whose directions are prescribed by the crystal symmetry determine the metastable spin textures~\cite{gobel2018magnetic}. For all types of DMI the energy contributions of the two skyrmions with reversed in-plane magnetizations --- what is necessary for the formation of \edb{the here considered} biskyrmion --- cancel. As a result, $H_\mathrm{DMI}$ is independent of $\Delta r$ and does not allow for the formation of a biskyrmion. Even more detrimental, at least one of the two skyrmions would annihilate in a simulation.

\begin{figure*}[t!]
	\centering
	\includegraphics[width=\textwidth]{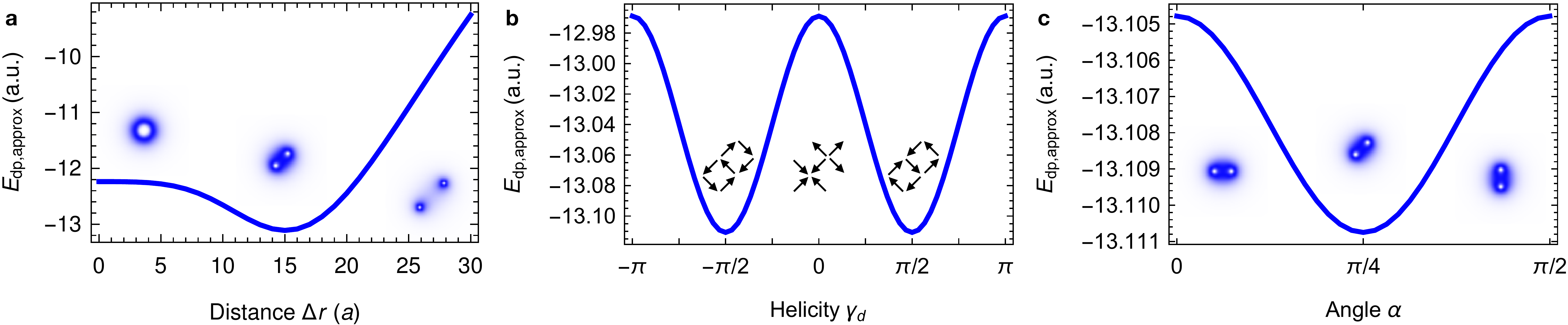}
	\caption{Formation of a biskyrmion by the dipole-dipole interaction. (\textbf{a}) Approximated dipole-dipole interaction energy [Eq.~\eqref{eq:dipole}] versus distance of the two skyrmions in the model [Eq.~\eqref{eq:model}]. Insets visualize the energy density for an $N_\mathrm{Sk}=2$ skyrmion with $\Delta r = 0$ (left), a biskyrmion with $\Delta r = 15 \, a$ (center), and for two skyrmions with $\Delta r = 30 \, a$ (right); $a$ is the lattice constant. Here, $\alpha = \pi/4$ and $\gamma_d=\pi/2$. The size parameter is $r_0=8a$ in all panels. (\textbf{b}) Helicity dependence of the approximated dipole-dipole energy for the biskyrmion with $\Delta r= 15 \, a$ and $\alpha = \pi/4$. Insets show the in-plane magnetization, representing two superposed Bloch skyrmions for $\gamma_d=\pm \pi/2$ and N\'{e}el skyrmions for $\gamma_d=0$. In all cases the two skyrmions have helicities $\gamma_d$ and $\gamma_d+\pi$, respectively. (\textbf{c}) Approximated dipole-dipole energy for the biskyrmion with $\Delta r = 15 \, a$ and $\gamma_d = \pi/2$ versus the biskyrmion orientation angle $\alpha$. Insets visualize the energy density.}
	\label{fig:energy}
\end{figure*}

\edc{An interaction that is present in all magnetic materials, but `overshadowed' by the DMI in non-centrosymmetric materials, is the dipole-dipole interaction
\begin{align}
H_\mathrm{dp} & = -\frac{1}{2}\sum_{ij} \frac{\mu_0}{4\pi |\vec{r}_{ij}|^5}[3(\vec{m}_i\cdot \vec{r}_{ij})(\vec{m}_j\cdot\vec{r}_{ij})-\vec{m}_i\cdot\vec{m}_j].\label{eq:dp}
\end{align} 
In case the DMI is forbidden, like in centrosymmetric materials, this interaction can stabilize not only skyrmions~\cite{lin1973bubble,takao1983study,jiang2015blowing,buttner2018theory} but also biskyrmions, as will be presented later in this paper by micromagnetic simulations. To establish an intuitive understanding of the attractive skyrmion-skyrmion interaction, which is necessary for the formation of a biskyrmion, we consider a short-range approximation of the dipole-dipole interaction for the calculations in this section. Note however, that this approximation is accompanied by a loss of information~\cite{aharoni2000introduction}: Even the lowest energy state of the considered model would relax to a collinear ferromagnetic configuration if only the short-range approximation was considered in a simulation in which shape relaxation is allowed. Since Derrick's theorem~\cite{derrick1964comments} implies the absence of local energy minima for non-collinear spin textures for this reduced type of dipole-dipole interaction~\cite{rajaraman1982solitons}, we cannot make statements on the metastability of a certain configuration. However, comparing the energies of states described by the model in Eq. \eqref{eq:model} allows to extract an attractive skyrmion-skyrmion interaction. The actual proof of metastability are the micromagnetic simulations of the three-dimensional system with the full dipole-dipole interaction shown later in this paper. A three-dimensional consideration is essential since non-collinear spin textures are only metastable once the sample is thicker than a critical value~\cite{gioia1997micromagnetics,eschenfelder2012magnetic}.} 

\edc{To extract the attractive interaction we consider in this section only nearest-neighbor interactions in a two-dimensional square lattice.} The second term in Eq. \eqref{eq:dp} merely rescales the exchange interaction, thereby rendering only the first term relevant
\begin{align}
H_\mathrm{dp,approx}\propto-\sum_{\braket{ij}}(\vec{m}_i\cdot \hat{\vec{r}}_{ij})(\vec{m}_j\cdot\hat{\vec{r}}_{ij}),\label{eq:dipole}
\end{align}
where $\hat{\vec{r}}_{ij}$ is a unit vector which is parallel to either $\hat{\vec{x}}$ or $\hat{\vec{y}}$. This interaction can be interpreted as a special easy-plane anisotropy. It favors parallel spins energetically. \edc{Such an alignment is approximately fulfilled} for the overlap region of the two skyrmions (that is the center of the biskyrmion). As a result, two skyrmions in centrosymmetric materials attract each other.

We find that partially overlapping Bloch skyrmions ($\gamma=\pm\pi/2$) at an displacement angle of $\alpha=\pm\pi/4$ or $\pm 3\pi/4$ have the lowest energy \edc{of all states described by Eq. \eqref{eq:model} with respect to the short-range approximation of the dipole-dipole interaction}. The energy minimum occurs at an inter-skyrmion distance of $\Delta r \approx 2 \, r_0$, which is a biskyrmion state (Fig.~\ref{fig:energy}a). The helicity dependence (Fig.~\ref{fig:energy}b) has a cosine shape making the Bloch skyrmion energetically most favorable. This helicity dependence is not unique to the biskyrmion state; a similar dependence is found for a single skyrmion. The $\alpha$ dependence (Fig.~\ref{fig:energy}c; global rotation of the texture) originates from the underlying lattice. As a consequence it has a period of $\pi/2$ with respect to $\alpha$ for the square lattice. \ed{Different lattices have energy minima at other angles, e.\,g. $\alpha = \pm \pi/6, \pm 3\pi/6$ and $\pm 5\pi/6$ for a hexagonal lattice.}

So far, we have shown that \edc{two skyrmions with reversed in-plane magnetizations attract each other} in centrosymmetric systems without DMI but with dipole-dipole interaction. The resulting biskyrmion is expected to be of Bloch type and to form at an angle of $45^\circ$ with respect to the underlying lattice. \edc{In the following, micromagnetic simulations will be used to analyze how these two specific skyrmions can be stabilized in a three-dimensional sample considering the full dipole-dipole interaction. We will show that all three results from the analytic model are confirmed in the simulation: the skyrmions attract each other to form a biskyrmion characterized by $\gamma_e=\pm \pi/2$ and $\alpha=\pm\pi/4,\pm 3\pi/4$.}

\subsection*{Setup for the micromagnetic simulations}
To support the above analytical considerations, we performed micromagnetic simulations for a thin magnetic nanodisk with out-of-plane anisotropy. \ed{These disks have been in the focus of skyrmion-related research in many recent publications~\cite{guslienko2015skyrmion, vidal2017stability, riveros2018analytical, guslienko2018neel, tejo2018distinct, allende2018skyrmion}.} The considered disk and its surfaces exhibit negligibly small DMI, so that only the dipole-dipole interaction (included in the demagnetization field) could stabilize a biskyrmion. 

The micromagnetic simulations were performed using the Mumax3 code~\cite{vansteenkiste2011mumax,vansteenkiste2014design}. Therein, the Landau-Lifshitz-Gilbert equation (LLG)~\cite{landau1935theory,gilbert2004phenomenological,slonczewski1996current} with parameters similar to those in Ref.~\cite{buttner2018theory} is solved for each discretized magnetic moment $\vec{m}_i$; see Methods.

Initially, two energetically degenerate metastable skyrmions with helicities $\pm \pi/2$ are `written' symmetrically to the nanodisk's center. Upon relaxation they are pushed to the center by their attractive interaction and the confinement potential of the nanodisk. When the skyrmions reach the energetically optimal distance, a biskyrmion has formed. We proceed by discussing the individual steps in detail and begin with the mechanism for writing the skyrmions.

\subsection*{Generation of isolated skyrmions}
The vanishing DMI complicates a controlled writing of skyrmions in centrosymmetric materials, since there is no energetically preferred chirality: injecting spins oriented along $\vec{z}$ via spin torque, as commonly done in DMI-dominated systems~\cite{sampaio2013nucleation}, produces a topologically trivial bubble (the in-plane spin components point toward the disk's center; they are oriented nearly parallel). Evidently, in order to write two skyrmions off-center without DMI the aimed-for texture has to be induced explicitly. In other words, deterministic instead of stochastic writing has to be considered. This is especially important since the two skyrmions need to have opposite helicities. 

\begin{figure*}[t!]
	\centering
	\includegraphics[width=\textwidth]{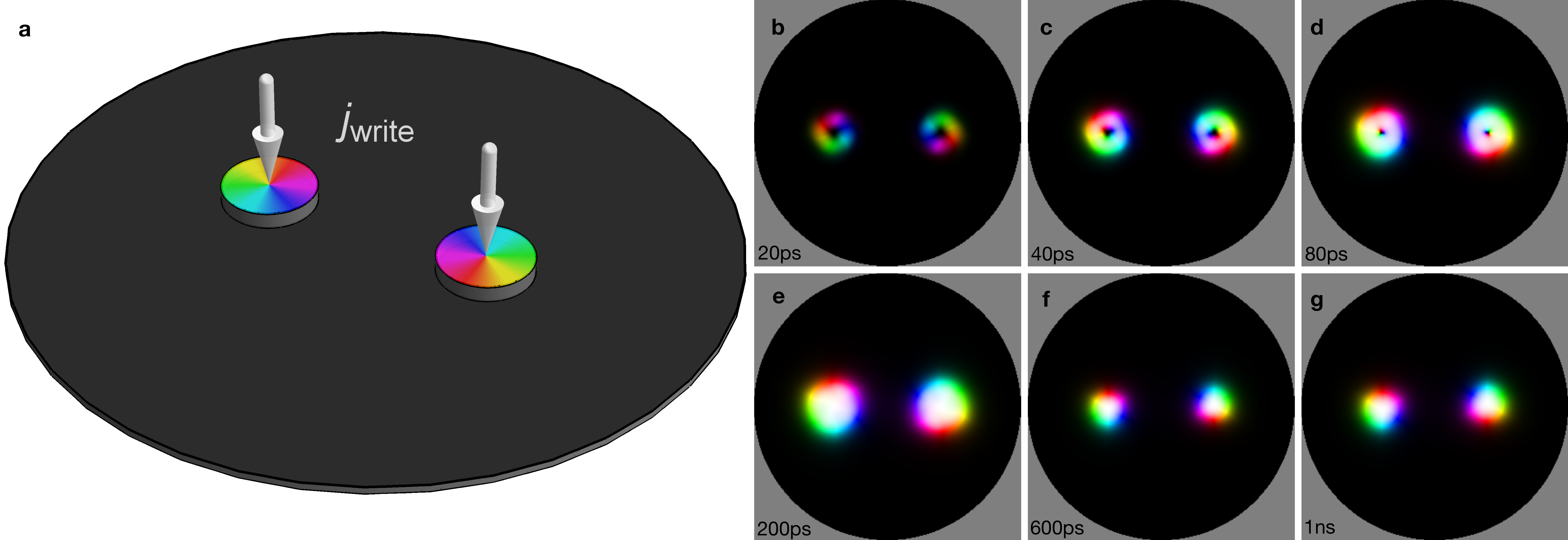}
	\caption{Writing two Bloch skyrmions with opposite helicities into a nanodisk. (\textbf{a}) Thin nanodisk with two smaller writing disks that host vortex textures (colored) with opposite helicities. The writing process is triggered by spin-polarized current densities $\vec{j}_\mathrm{write}$ which are applied for $200\,\mathrm{ps}$. (\textbf{b}--\textbf{g}) Snapshots during the generation of two skyrmions. The color scale is the same as in Fig.~\ref{fig:overview}. An animated version is accessible in Supplementary Video 1.}
	\label{fig:disk}
\end{figure*}

We consider two smaller disks attached to the nanodisk (Fig.~\ref{fig:disk}a). These additional disks host stable in-plane vortex configurations~\cite{cowburn1999single,shinjo2000magnetic}, similar to Refs.~\cite{sun2013creating, miao2014experimental,gilbert2015realization,loreto2018creation}. The opposite helicities of the vortices, which are switchable e.\,g., by a magnetic field~\cite{taniuchi2005vortex,gaididei2008controllable,konoto2008formation, yakata2010control}, determine the helicity of the two emerging Bloch skyrmions. If an electric current $\vec{j}_\mathrm{write}$ is applied in perpendicular direction (here: $\vec{j}_\mathrm{write} =7\times 10^8 \mathrm{A}/\mathrm{cm}^{-2}\vec{e}_z$ applied for $200\,\mathrm{ps}$), the spin of the flowing electrons is aligned with the vortex (as long as the coupling is sufficiently large) and injected into the underlying nanodisk; there a skyrmion is formed via spin torque.

The writing process is visualized in Figs.~\ref{fig:disk}b--g. Starting from a ferromagnetic state, a skyrmionium-like configuration (a skyrmion with reversed skyrmion-like magnetization in its center~\cite{bogdanov1999stability,zhang2016control,gobel2019electrical}) is written (b--d). Since the `central' reversed skyrmion is very small, it quickly annihilates ($< 100\,\mathrm{ps}$) so that a skyrmion with a topological charge of $N_\mathrm{Sk}=+1$ remains at each side of the disk (e). On the time scale of $1\,\mathrm{ns}$ their sizes relax (f, g).

While the writing mechanism considered here appears convenient, other approaches are suited for an experimental realization as well; in the following, we give an incomplete list.

Merons (vortices with out-of-plane spins in their center) in the writing devices work equally well. The same holds for skyrmions. All of the considered textures have a similar in-plane magnetization which seems to be the most important component for skyrmion generation in centrosymmetric materials; due to the fact that the dipole-dipole interaction is achiral, a suitable in-plane magnetization has to be `imprinted' in order to `generate' a topological charge.

In a recent publication~\cite{gobel2019electrical} we have presented a mechanism for writing skyrmioniums via a photosensitive switch~\cite{yang2017ultrafast}: a laser triggers a radial current in a ring of a heavy metal, thereby generating an in-plane toroidal spin polarization via the spin Hall effect. This toroidal spin profile is then injected into the magnetic nanodisk, very similar to the vortex-writing device but for a ring geometry instead of a disk geometry. Depending on the chosen parameters, this method works for writing skyrmions as well. The sign of the applied bias voltage controls the spin polarization's orientation and thus the writing of either $+\pi/2$ or $-\pi/2$ Bloch skyrmions.

As a third option, one may consider spatiotemporally tuned electron sources~\cite{schaffer2017ultrafast}. The type of induced Bloch skyrmion can be controlled via the propagation direction of the electrons (beam applied from above or below the nanodisk).

\subsection*{Motion of a single skyrmion}
\begin{figure*}[t!]
  \centering
  \includegraphics[width=\textwidth]{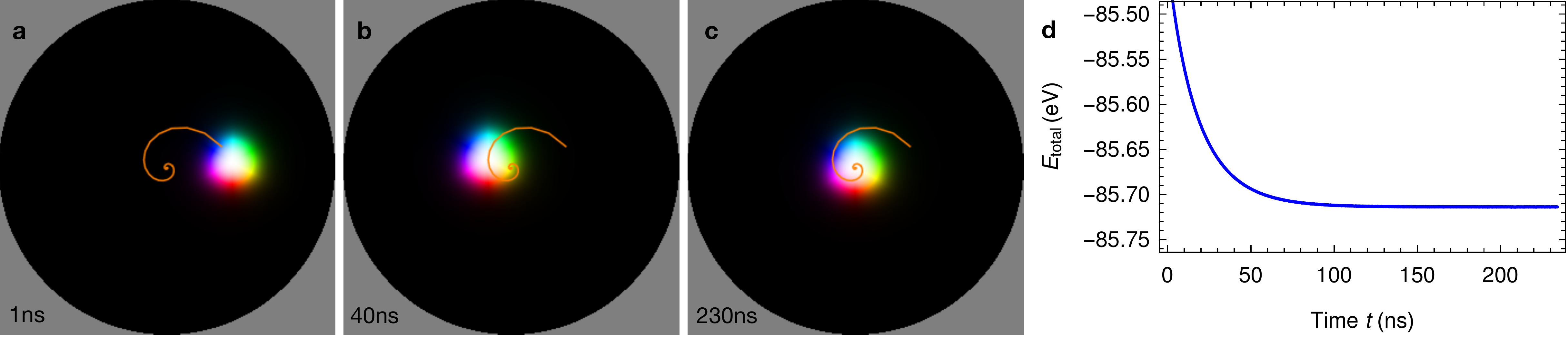}
  \caption{Motion of a single skyrmion within a nanodisk. (\textbf{a}--\textbf{c}) Snapshots and trajectory (orange line) of a single skyrmion in the nanodisk at indicated times. (\textbf{d}) Temporal progress of the system's total energy. An animated version is accessible in Supplementary Video 2.}
\label{fig:1skyrmion}
\end{figure*}

Before discussing the formation of biskyrmions, we address the motion of a single skyrmion within the nanodisk. After relaxation of shape and size, a skyrmion starts to move toward the center of the nanodisk in a counter-clockwise spiral trajectory (orange in Figs.~\ref{fig:1skyrmion}a--c). Initially, the skyrmion is accelerated but is subsequently decelerated when the skyrmion's center is close to the nanodisk's center. The helicity of the skyrmion remains unchanged during the evolution.

The spiral trajectory is understood by considering the Thiele equation~\cite{thiele1973steady,gobel2019overcoming} \begin{align}
b\,\vec{G} \times \vec{v} - b \underline{D} \alpha\vec{v} + F_\mathrm{int}(r) \frac{\vec{r}}{r} = 0\label{eq:thielesot}.
\end{align}
In this effective `center-of-mass-like' description of magnetic quasiparticles with velocity $\vec{v}$ all internal degrees of freedom have been integrated out. The spatial details of the skyrmion are `condensed' in the gyromagnetic coupling vector $\vec{G} = -4\pi N_\mathrm{Sk}\vec{e}_z$ and in the dissipative tensor $\underline{D}$ with elements $D_{ij} = \int\partial_{i}\vec{s}(\vec{r})\cdot\partial_{j}\vec{s}(\vec{r})\,\mathrm{d}^2r$. Only  $D_{xx} = D_{yy}$ are nonzero. $b=M_sd_z/\gamma_e$ is determined by the saturation magnetization $M_s$, the disk thickness $d_z$ and the gyromagnetic ratio of an electron $\gamma_e$. Due to the disk geometry the force $F_\mathrm{int}$, that covers interactions of the considered quasiparticle with other quasiparticles or with the edge of the sample, is radially symmetric. The Thiele equation in polar coordinates
\begin{align}
4\pi N_\mathrm{Sk}b(r\dot{\phi}\vec{e}_r-\dot{r}\vec{e}_\phi)-\alpha D_{xx}b(\dot{r}\vec{e}_r+r\dot{\phi}\vec{e}_\phi)+F_\mathrm{int}(r)\vec{e}_r = 0
\end{align}
 yields 
\begin{align}
\label{eq:eom}
\begin{split}
\dot{r}&=\frac{F_\mathrm{int}(r)}{(\alpha D_{xx})^2b+(4\pi N_\mathrm{Sk})^2b}\alpha D_{xx},\\
\dot{\phi}&=\frac{F_\mathrm{int}(r)}{(\alpha D_{xx})^2b+(4\pi N_\mathrm{Sk})^2b}\frac{-4\pi N_\mathrm{Sk}}{r}.
\end{split}
\end{align}
Since $F_\mathrm{int}(r) < 0$, a skyrmion moves to the nanodisk's center along a  counter-clockwise spiral trajectory ($\dot{r}<0$, $\dot{\phi}>0$).  When it approaches the center, the angular velocity decreases since $F_\mathrm{int}(r)$ drops faster to $0$ than $1/r$. The transverse component of the spiral motion is a consequence of the skyrmion's topological charge. Like any quasiparticle with nontrivial real-space topology a skyrmion does not move parallel to the propelling force --- let it be due to a current or the skyrmion--edge interaction~\cite{nagaosa2013topological,zang2011dynamics,jiang2017direct,litzius2017skyrmion, gobel2019overcoming,gobel2018magnetic} --- and thus exhibits a skyrmion Hall effect. The corresponding transverse force is similar to the Magnus force for classical particles~\cite{everschor2014real,nagaosa2013topological}.

\subsection*{Formation of biskyrmions}
The formation of a biskyrmion starts with two Bloch skyrmions with opposite helicities of $\gamma = \pm \pi/2$, as stabilized in Fig.~\ref{fig:disk}. Since both skyrmions have identical dissipative tensors $\underline{D}$ and topological charges $N_\mathrm{Sk}=+1$, the equations of motion~\eqref{eq:eom} hold for both quasiparticles. They move on a spiral trajectory towards the nanodisk's center. However, the skyrmion-skyrmion interaction has to be considered in $\vec{F}_\mathrm{int}$. This force changes sign when both skyrmions come close to each other, corresponding to the energy minimum in Fig.~\ref{fig:energy}a; it is zero at a finite distance of the two skyrmions.

\begin{figure*}[t!]
  \centering
  \includegraphics[width=\textwidth]{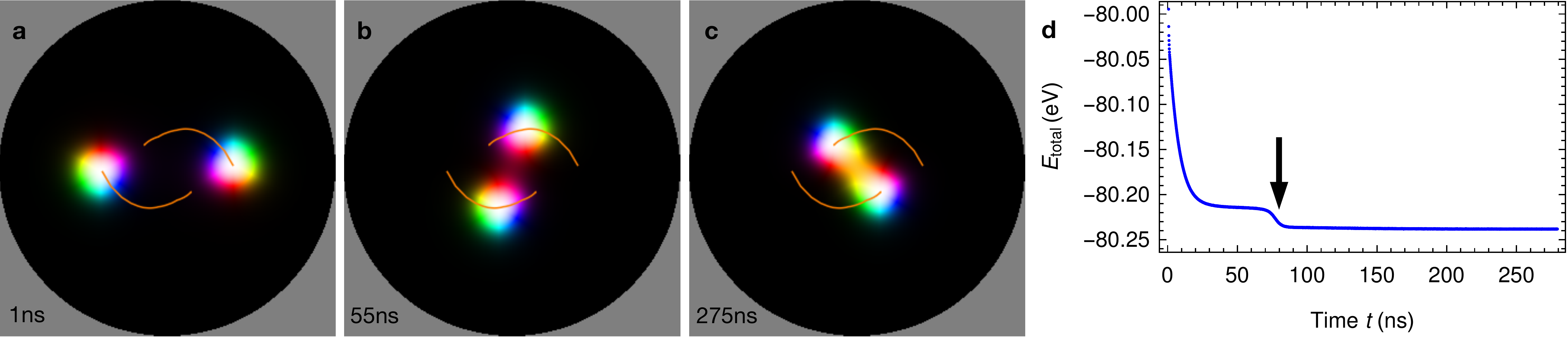}
  \caption{Formation of a biskyrmion. (\textbf{a}--\textbf{c}) Snapshots and trajectories (orange) of two Bloch skyrmions with opposite helicities. (\textbf{d}) Temporal progress of the system's total energy. The pronounced drop (indicated by an arrow) corresponds to the energy minimum in Fig.~\ref{fig:energy}a: a stable biskyrmion is formed at that point in time. An animated version is accessible in Supplementary Video 3.}
\label{fig:biskyrmion}
\end{figure*}

This behavior is well visible in the results of micromagnetic simulations (Fig.~\ref{fig:biskyrmion}). After the writing period (relaxation after $\approx 1\,\mathrm{ns}$ as presented in Fig.~\ref{fig:disk}), both skyrmions start to exhibit a spiral motion to the center as in the case of a single skyrmion. Once both skyrmions start to overlap the spiral motion stops and a biskyrmion is formed. During the formation process the topological charge remains $N_\mathrm{Sk}=+2$ \ed{within an error of less than $\unit[1.5]{\%}$ (the topological charge is not integer due to the discreteness of the underlying lattice)}.

The biskyrmion configuration is remarkably stable. The energy drops significantly once the two skyrmions are close to each other (Fig.~\ref{fig:biskyrmion}d). This drop corresponds to the energy minimum established in the analytic model (Fig.~\ref{fig:energy}a). In the final phase of the propagation the biskyrmion aligns with one of the crystallographic directions $\alpha = 45^\circ$, $135^\circ$, $225^\circ$ or $315^\circ$, visible as energy minimum in Fig.~\ref{fig:energy}c (here: $\alpha = 135^\circ$). Once the system is close to the energy minimum $\vec{F}_\mathrm{int}$ is far from being rotational symmetric and the equations of motion derived from the Thiele equation do not hold anymore. By comparing the energy scales of Figs.~\ref{fig:energy}a and~\ref{fig:energy}c it becomes apparent that the reorientation of the biskyrmion plays a role after the optimal distance of the two skyrmions has been reached.
The distinguished axis of the biskyrmion can in principle be imposed by the initial positions of the skyrmions. If the two writing disks are closer together the biskyrmion can end for example at $\alpha = 45^\circ$ (Fig.~S1 of the Supplementary information).

Starting with two skyrmions with \emph{identical} helicity (i.\,e.\ by switching one of the two writing vortices) a mostly static behavior is observed (Fig.~S2 of the Supplementary information). This is readily explained by the stronger repulsion of skyrmions with the same helicity in comparison to those with opposite helicities. Due to the disk geometry both skyrmions experience a force towards the center, resulting in a zero of $F_\mathrm{int}$ at larger distances compared to the biskyrmion system. The (final) steady state --- two skyrmions without considerable overlap ---  has a sizably higher energy than the biskyrmion state (cf. Figs.~S2d and Fig.~\ref{fig:biskyrmion}d). Starting from this particular configuration a biskyrmion does not form since one skyrmion would have to reverse its helicity. That, however, requires to overcome an energy barrier similar to that in Fig.~\ref{fig:energy}b, corresponding to a N\'{e}el skyrmion state.

\edb{In Refs.~\cite{leonov2016three,du2018interaction} it was shown that asymmetric skyrmion tubes (e.\,g. in the cone phase of chiral magnetic materials) can exhibit an attractive interaction despite their equal helicities. The resulting non-overlapping pair of skyrmions is stabilized by a different mechanism and at a larger inter-skyrmion distance compared to the type of biskyrmion discussed in the present paper. Nevertheless, the skyrmion-skyrmion pair was labeled `biskyrmion' in Ref.~\cite{leonov2016three} due to the topological equivalence ($N_\mathrm{Sk}=\pm 2$) of both textures. Yet, both are geometrically distinct objects.}

\section*{Discussion}
Here, we have demonstrated that the dipole-dipole interaction in centrosymmetric magnetic systems can stabilize \emph{isolated biskyrmions}\edb{, in the sense of a pair of partially overlapping skyrmions with reversed in-plane magnetizations}. We propose methods of writing two Bloch skyrmions with opposite helicities at opposite sides of a nanodisk. After spiral propagation of each skyrmion towards the nanodisk's center, an \emph{isolated biskyrmion} is formed due to the attractive interaction that originates in the dipole-dipole interaction. These ingredients provide a step towards utilizing biskyrmions as carriers of information in spintronics devices. 

\ed{To the best of our knowledge, only biskyrmion \emph{lattices} have been found experimentally so far~\cite{yu2014biskyrmion,wang2016centrosymmetric, peng2017real,zuo2018direct,peng2018multiple}, and even for them the interpretation of the presented Lorentz transmission electron microscopy images is under serious debate. In two recent publications it was shown that tubes of topologically trivial bubbles (called type-II or hard bubbles) can appear as biskyrmion-like features when observed under an angle~\cite{loudon2019images,yao2019magnetic}. Since thin magnetic disks are considered in the present work, potential real-space images can hardly be misinterpreted, and therefore our prediction may be decisive for showing that biskyrmions can exist at all.}

Besides their potential for applications, biskyrmions are also worth being investigated from a fundamental point of view. Their nontrivial real-space topology, manifested in the topological charge of $N_\mathrm{Sk} = \pm 2$, imposes emergent electrodynamic effects: just like conventional skyrmions, biskyrmions exhibit a skyrmion Hall effect~\cite{nagaosa2013topological,zang2011dynamics,jiang2017direct,litzius2017skyrmion, gobel2019overcoming} and a topological Hall effect~\cite{bruno2004topological,neubauer2009topological,hamamoto2015quantized, gobel2017THEskyrmion,gobel2017QHE,gobel2018family, hamamoto2016purely,maccariello2018electrical,nagaosa2013topological,nakazawa2018topological}. In other words, under the effect of a spin-polarized current the biskyrmion and the current electrons are deflected into transverse directions. Our study motivates future in-depth analyses of similarities and differences of these Hall effects for skyrmions and biskyrmions.

One fundamental aspect that distinguishes biskyrmions from skyrmions is their missing rotational symmetry. The orientation of a biskyrmion could be exploited to store multiple bits per biskyrmion. Here, the underlying cubic lattice yields four energetically degenerate orientations $\alpha$ and allows for quaternary instead of binary logic. It is conceivable to store data in an array of nanodisks instead in a racetrack device.

\section*{Methods}
For the micromagnetic simulations the Landau-Lifshitz-Gilbert equation~\cite{landau1935theory,gilbert2004phenomenological, slonczewski1996current}
\begin{align}
\dot{\vec{m}}_i=-\gamma_e\vec{m}_i\times\vec{B}_{i,\mathrm{eff}}+\alpha_g\vec{m}_i\times\dot{\vec{m}_i}+\gamma_e \epsilon\beta[(\vec{m}_i\times\vec{s}_i)\times\vec{m}_i]
\end{align}
is solved using Mumax3~\cite{vansteenkiste2011mumax,vansteenkiste2014design}.

The first term describes the precession of each magnetic moment around its instantaneous  effective magnetic field $\vec{B}_{\mathrm{eff}}^i = -\delta_{\vec{m}_i} F /M_s$. This field, derived from the free energy density $F$, takes into account exchange interaction, uniaxial anisotropy, Zeeman energy, and the demagnetization field accounting for the dipole-dipole interaction. 

The second term, introduced phenomenologically, considers that the magnetic moments tend to align along their effective magnetic field. The strength of this damping is quantified by the Gilbert damping constant $\alpha_g$. 

The third term accounts for the coupling of the magnetic moments with injected spin-polarized currents. The in-plane torque coefficient $\epsilon \beta = \hbar P j / 2ed_z M_s$ depends on the current's spin polarization $P$ and the current density $j$. The spatial distribution of the injected spins $\{\vec{s}_i\}$ is given by the vortex textures in the writing disks. An  out-of-plane torque has been neglected since it is usually small and merely rescales the external magnetic field. $\gamma_e = 1.760 \times 10^{11}\mathrm{T}^{-1} \mathrm{s}^{-1}$ is the gyromagnetic ratio of an electron.

We use parameters that stabilize stray-field skyrmions, similar to those in Ref.~\cite{buttner2018theory}: exchange $A = 15\,\mathrm{pJ}/\mathrm{m}$, uniaxial anisotropy in $z$ direction $K_z = 1.2\,\mathrm{MJ} / \mathrm{m}^3$, external field $\vec{B} = -40\,\mathrm{mT} \,\vec{e}_z$ (along the ferromagnetic orientation), saturation magnetization $M_s = 1.4\,\mathrm{MA} / \mathrm{m}$, and Gilbert damping $\alpha_g = 0.3$. The disk has a radius of $150\,\mathrm{nm}$ and a thickness of $d_z=3\,\mathrm{nm}$. For the simulations we used discrete magnetization cells of size $1\,\mathrm{nm} \times 1\,\mathrm{nm} \times 1\,\mathrm{nm}$. \ed{The DMI has been set to zero. We checked that a small DMI constant (interfacial DMI $D = 0.1\,\mathrm{mJ}/\mathrm{m}^2$) does not qualitatively affect the results of the simulations shown in Fig.~\ref{fig:biskyrmion}; for larger values of $D$ one of the two skyrmions would annihilate and a biskyrmion would not form.}

As stated in the main text, a spin-polarized current of $\vec{j}_\mathrm{write}P = 7\times 10^8 \mathrm{A}/\mathrm{cm}^{-2}\vec{e}_z$ with a spin polarization orientation $\{\vec{s}_i\}$, given by an in-plane vortex texture
\begin{align}
\vec{s}_i=\pm(-y,x\mp\Delta x/2,0)/\sqrt{(x_i\pm\Delta x/2)^2+y^2},
\end{align}
writes well-defined skyrmions into the nanodisk, if the current is applied for $200\,\mathrm{ps}$. The two signs correspond to the left and right writing disks, respectively. The writing disks have a radius of $25\,\mathrm{nm}$ each and their centers are at a distance of $120\,\mathrm{nm}$. \ed{The writing disks and the process of spin polarization have not been simulated explicitly but are condensed into the spin-polarization parameter $P$ which scales the necessary applied current density.}

\bibliography{sample}

\section*{Acknowledgements}
B.G. is thankful to Alexander F. Sch{\"a}ffer regarding discussions about Mumax and possible alternative writing approaches. This work is supported by Priority Program SPP 1666 and SFB 762 of Deutsche Forschungsgemeinschaft (DFG).

\section*{Author contributions statement}
B.G. initiated the project and performed the calculations. B.G. wrote the manuscript with the help of J.H. All authors discussed the results and commented on the manuscript.

\section*{Additional information}
\subsection*{Supplementary information}
Supplementary information accompanies this paper at [URL].

\subsection*{Accession codes}
The code Mumax3 is accessible at https://github.com/mumax/3/releases/tag/v3.9.3.

\subsection*{Competing interests}
The authors declare no competing interests.

\end{document}